\documentclass[prb,twocolumn,showpacs]{revtex4}
\usepackage{graphicx}
\usepackage{epsfig}
\usepackage{longtable}
\newcommand{\is}{\it {s}}
\newcommand{\ih}{\it {h}}
\newcommand{\ip}{\it {p}}
\newcommand{\ic}{\it {c}}
\newcommand{\ib}{\it {b}}
\newcommand{\bb}{\it {b-b}}
\newcommand{\kk}{\it {s-s}}
\newcommand{\ks}{$s$-$s^\prime$}
\newcommand{\pp}{\it {p-p}}
\newcommand{\cc}{\it {c-c}}
\newcommand{\hh}{\it {h-h}}
\newcommand{\ph}{\it {p-h}}

\begin{document}

\title{A study of Mg adsorption on Si(001) surface from first principles}

\author{R.~Shaltaf, E.~Mete, and \c{S}.~Ellialt{\i}o\u{g}lu\footnote{Corresponding
author. E-mail: sinasi@metu.edu.tr}}

\affiliation{Department of Physics, Middle East Technical University, Ankara
06531, Turkey}

\date{\today}

\begin{abstract}

First-principles calculations using density functional theory based on
norm-conserving pseudopotentials have been performed to investigate the Mg
adsorption on the Si(001) surface for 1/4, 1/2 and 1 monolayer coverages. For
both 1/4 and 1/2 ML coverages it has been found that the most favorable site
for the Mg adsorption is the cave site between two dimer rows consistent with
the recent experiments. For the 1 ML coverage we have found that the most
preferable configuration is when both Mg atoms on 2$\times$1 reconstruction
occupy the two shallow sites. We have found that the minimum energy
configurations for 1/4 ML coverage is a 2$\times$2 reconstruction while for the
1/2 and 1 ML coverages they are 2$\times$1.
\end{abstract}




\maketitle

\section{Introduction}

Adsorption of overlayers of Mg atoms on silicon surfaces has been a subject of
growing interest during the last decade because of its importance in
technological applications such as the efficient photocathodes and thermionic
energy converters. Even though several experimental and theoretical studies
have been carried out for studying the Mg/Si(111) systems\cite{s111-theor1}
only few have been reported for the adsorption of Mg on Si(001) surfaces.
\cite{BGK,Kaw,HEN,Kim,Kubo,Cho,KO} The understanding of Mg adsorption on
Si(001) has become especially an interesting problem to investigate because of
the possible use of Si substrates for the growth of $\rm Mg_2Si$ films. More
recently, after the report of superconductivity in $\rm MgB_2$, its growth on
Si(001) surfaces\cite{Bri,Ple} has become a current issue.

One of the earliest experiments of Mg adsorption on Si(001) were carried out by
Kawashima et al.,\cite{Kaw} who have obtained LEED and AES results starting
from full coverage to lower coverages by allowing for thermal desorption. They
observed that as the coverage decreased the various structural phases appeared
in the order 1$\times$1, 2$\times$3, 2$\times$2, another 2$\times$3, and
finally the clean 2$\times$1 which correspond to the coverages of 1, 1/3, 1/4,
1/6, and 0 ML, respectively. They concluded that Mg atoms were adsorbed on
hollow site (valley-bridge site, in their notation) for each coverage mentioned
above.

Hutchison et al.,\cite{HEN} in their STM results for the low
coverage-Mg/Si(001) case, have reported three types of adsorption geometries at
room temperature. Type I is the most favorable phase and it refers to a single
Mg atom adsorbed on a cave site. The least frequently observed type II, being a
localized phase, corresponds to two Mg atoms adsorbed on cave sites to relieve
the stress on the chain of type I phases. Type III phase, which may occur
everywhere but next to a type II phase, contains multiple Mg adatoms (three at
most) adsorbed on a combination of adjacent sites.

Similar observations also with the use of STM have been reported by Kubo et
al.\cite{Kubo} for Mg adsorption at room temperature. However, they speculated
another interpretation differing from that of Hutchison et al., such that
according to Kubo et al., type I may correspond to two Mg atoms adsorbed on
adjacent hollow sites and type III may correspond to a single Mg atom adsorbed
on a pedestal site. In addition, for the annealed case at high temperature,
they have tentatively suggested two models possessing the 2$\times$2 symmetry
by incorporating an extra Si on the shallow site (see Fig 1) bonding to Mg
adatom. The first model consists of a single Mg atom adsorbed on the
neighboring shallow site bonding across the hallow site, and second one
consists of two Mg atoms adsorbed on cave sites on each side of the extra Si
atom.

Kim et al.\cite{Kim} have reported in their LEED data that the 2$\times$2
reconstruction corresponding to 1/2 ML coverage occurs at 280$^\circ$C, whereas
for 1/3 ML coverage with 2$\times$3 symmetry occurs at 390$^\circ$C showing the
dependence of surface reconstruction on coverage. Kubo et al.\cite{Kubo} did
not reject the possibility of 1/2 ML as saturation coverage for 2$\times$2
surface, either. Cho et al.\cite{Cho} have performed high resolution core-level
photoelectron spectroscopy for 2$\times$2 and 2$\times$3 structures. They did
not find any difference in the behavior of Mg adsorption on Si(001) surface for
these two configurations. They have found that the most preferable adsorption
site for both of these low coverages is the cave site between two dimers
(bridge site, in their notation).

On the theory side, Khoo and Ong\cite{KO} using a semiempirical self-consistent
molecular orbital method have performed CNDO (complete neglect of differential
overlap) calculations to investigate the adsorption of Mg atom on Si(001) at
1/2 ML coverage. According to their results the Mg atom resides on the bridge
site above the midpoint of Si dimer.

Wang et al.,\cite{Wang} have done first principle calculations for the
adsorption of another group II element Ba on Si(001) at a rather low coverage
of 1/16 ML, and they have shown that the minimum energy site is the hollow site
(cave-bridge site, in their notation) where the Ba atom is surrounded by four
buckled dimers leading to a solitonlike defect in the original $c(4\times2)$
configuration.

The aim of this paper is to perform first-principle calculations for different
coverages, i.e., 1/4, 1/2 and 1 ML, of Mg on Si(001) to get a clear
understanding of the adsorption mechanisms and to investigate the atomic
structure of the surface covered with magnesium. To the best of our knowledge
this is the first detailed work to investigate theoretically the Mg/Si(001)
system for different coverages from first principles.

\section{Method}

We used pseudopotential method based on density functional theory in the local
density approximation. The self consistent norm conserving pseudopotentials are
generated by using the Hammann scheme\cite{Ham-89} which is included in the
fhi98PP package.\cite{FS-99} Plane waves are used as a basis set for the
electronic wave functions. In order to solve the Kohn-Sham equations, conjugate
gradients minimization method\cite{Pay} is employed as implemented by the
ABINIT code.\cite{abinit,Gonze} The exchange-correlation effects are taken into
account within the Perdew-Wang scheme\cite{PW-92} as parameterized by Ceperly
and Alder.\cite{CeperlyAlder}

The unit cell included an atomic slab with 8 layers of Si plus a vacuum region
equal to about 9 {\AA} in thickness. Single-particle wave functions were
expanded using a plane wave basis up to a kinetic energy cut-off equal to 16
Ry. The integration in the Brillioun zone was performed using 8 special $\vec
k$-points sampled using Monkhorst-Pack\cite{MP-76} scheme. Although a couple of
cases were repeated with 18 special $\vec k$-points, no significant improvement
has been observed.

We have used our theoretical equilibrium lattice constant for the bulk Si
(5.405 \AA) in the surface calculations. The bulk modulus for Si is found to be
96 Mb in rather good agreement with the experimental result of 97 Mb. Our
results for Mg bulk lattice parameter is 3.12 {\AA}, and the c/a ratio is 0.607
that are very close to the experimental results of 3.20 {\AA} and c/a of 0.614,
respectively. As another check for our pseudopotentials, the lattice parameter
for $\rm Mg_2Si$ was calculated to be 6.28 {\AA} which is also in close
agreement with the experimental value of 6.39 {\AA}.\cite{Wyck} The energy
bands of this compound gives a semiconducting gap of E$_{\rm g}$ equal to 0.31
eV. These tests suggest that the inclusion of the nonlinear core corrections to
Mg pseudopotential is not needed.

We have used the 2$\times$2 surface unit cell in our calculations to study the
adsorption of Mg at a low coverage such as 1/4 ML and to include various
combinations of adsorption sites for the half- and full-monolayer coverages.
The first step was to optimize the clean Si(001)-2$\times$2 surface while
keeping the two lowest substrate layers (out of 8) frozen into their bulk
positions and all the remaining substrate atoms were allowed to relax into
their minimum energy positions. The $p(2\times$2) was found to have the lowest
total energy with a dimer length of 2.32 {\AA} and a tilt angle of $19^\circ$
which is in good agreement with the experimental dimer-length value of 2.40
$\pm$ 0.10 \AA.

\section{Results and Discussion}

We have studied the adsorption of Mg atom on the Si(001) surface for 1/4, 1/2
and 1 ML starting with the reconstructed $p(2\times$2) surface unit cell. We
have chosen five different sites for adsorption, namely, cave, hallow,
pedestal, bridge and shallow. The cave site ({\ic}) is located above the fourth
layer Si, hallow ({\ih}) and pedestal ({\ip}) sites are above the third layer
Si, shallow site ({\is}) above the second layer Si between two Si dimers, and
the bridge site ({\ib}) is located above the dimer as indicated in Fig. 1. The
adatom and the upper three monolayers of the substrate were then taken to their
minimum energy configurations by performing structural optimization using the
Broyden-Fletcher-Goldfarb-Shanno method\cite{BFGS} until the force on each atom
reduces to a value less than 25 meV/\AA.

\begin{figure}[htb]
\includegraphics[width=8cm]{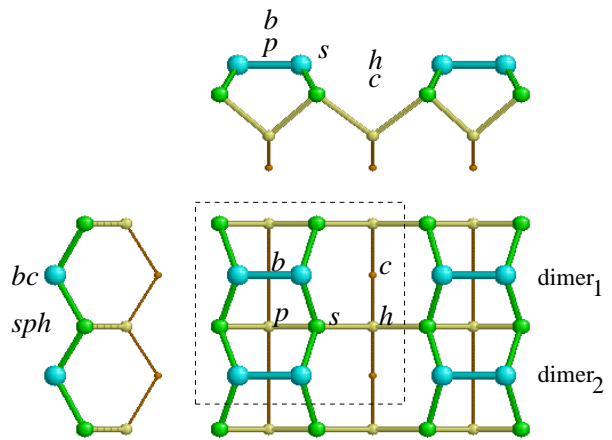}
\caption{Schematics of the adsorption sites for Mg atom on Si(001) surface from
three different views. The symbols stand for: $b$=bridge, $p$=pedestal,
$h$=hallow, $s$=shallow and $c$=cave. (The dimers are shown symmetric here for
visual convenience)\label{figure1}}
\end{figure}

In Table~\ref{table1} we introduce the structural parameters and bond lengths
for different adsorption sites as well as the adsorption energy $\rm E_{ad}$
(negative of the binding energy of the adatom) given by \begin{equation}
E_{ad}=E_{\rm Si(001)}+nE_{\rm Mg}-E_{\rm Mg/Si(001)} \end{equation} where
$E_{\rm Mg/Si(001)}$ is the total energy of the adsorbed surface, $E_{\rm
Si(001)}$ the total energy of clean surface, $n$ is the number of Mg adatoms in
the surface unit cell and $E_{\rm Mg}$ is the total energy of a single Mg atom
with spin polarization obtained in a separate ab initio calculation using the
same pseudopotential and the same energy cut off in a larger unit cell with a
size of about 15 \AA.

\subsection{1/4 ML Coverage}

\begin{figure}[htbp]
\epsfig{file=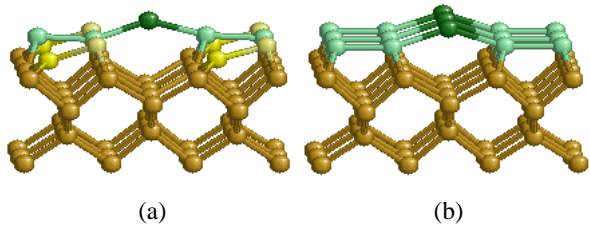,width=8cm,clip=true} \caption{Mg overlayer adsorbed on
Si(001) surface (a) for 1/4 ML coverage on {\ic} site and (b) for 1/2 ML
coverage on {\cc} site \label{figure2}}
\end{figure}

The adsorption of Mg atom on cave site was found to be the most favorable one
among the five different adsorption sites, shown in Fig. 1. The other cases,
namely, the {\is} site and the {\ib} site occupations by an adatom are also
stable. However, in the case of adsorption of Mg on {\ip} or {\ih} sites, the
adatoms reside there temporarily (for some optimization steps) before it
becomes unstable and then they migrate to different sites with lower energies.
In fact, the adatom on {\ip} site migrates to the {\is} site, while the adatom
on {\ih} site migrates to the neighboring {\ic} site.

The adsorption of Mg on {\ic} site corresponds to an adsorption energy of 2.26
eV and {\is} and {\ib} sites were found to be 0.25 eV and 0.68 eV lower than
that of {\ic} case, respectively. A natural pathway for surface diffusion might
be $b\rightarrow s\rightarrow c$. Listed in Table~\ref{table1} is also the
values for the metastable case of {\ih} site, since it takes longer for the
$h\rightarrow c$ migration, in contrary to the {\ip} site with much faster
$p\rightarrow s$ migration since the adatom there is in a very unstable
environment.

For all stable cases in 1/4 ML coverage we have found that the surface dimers
get partially symmetrized. Only one dimer (``dimer$_1$" to which Mg adatom
bonds) becomes symmetric (with a tilt angle less than $\sim6^\circ$), while the
other one (dimer$_2$ with no charge transfer from Mg adatom) is still tilted
(e.g., Fig. 2a). The tilting angles for dimer$_2$ in {\ic}, {\ib} and {\is}
cases being 16.0$^\circ$, 17.1$^\circ$ and 13.0$^\circ$, respectively, are
still close to the clean surface value of 19$^\circ$ while dimer$_1$ for all of
them are almost flat as seen from Table~\ref{table1}.

The height of Mg atom on cave site is found to be 0.75 {\AA} which is in good
agreement with the STM results of Hutchison et al.\cite{HEN} Fig. 2a shows for
the 1/4 ML coverage that the two Si atoms bonded to the same Mg adatom are
still in an asymmetric environment. The one whose neighbors are tilted down is
2.68 {\AA} away from Mg atom whereas the other has a bondlength of 2.65 {\AA}.
In a different calculation for $\rm Mg_2Si$ we have found Mg-Si bondlength to
be 2.70 {\AA} which is close to the back-bond lengths of Mg in this
environment.

\begin{table*}[htbp]
\caption{The structural parameters (in {\AA}) for most of the stable adsorption
sites (except for the starred case which is metastable) for three different
coverages $\Theta$. The heights of the adatom with respect to the dimers
(d$_\perp$ in \AA) and the adsorption energies ($\rm E_{ad}$ in eV) for these
adsorption cases are also presented. The quantities in parenthesis are tilt
angles (in degrees) of the corresponding dimers.\label{table1}}
\begin{minipage}{\textwidth}
\begin{tabular}{cc|cc|cc|cc|c|c}
~\\ \hline\hline &&&&&&&& \\[-1.5mm]
\hspace{5pt} $\Theta$ \hspace{5pt} & \hspace{10pt} Model \hspace{10pt} &
\hspace{10pt} dimer$_1$ \hspace{10pt} & \hspace{10pt} dimer$_2$ \hspace{10pt} &
\hspace{10pt} ${\rm d_{Mg-Si1}}$ \hspace{2pt} & \hspace{2pt} ${\rm d_{Mg-Si2}}$
\hspace{10pt} & \hspace{10pt} d$_{\perp1}$ \hspace{2pt} & \hspace{2pt}
d$_{\perp2}$ \hspace{10pt} & \hspace{5pt} ${\rm 2\times n}$
\hspace{5pt} & \hspace{10pt} E$_{\rm ad}$ \hspace{10pt} \\[1mm]
\hline\hline &&&&&&&&& \\[-1mm] 1/4 
& {\ic} & 2.54 (2.5) & 2.28 (16.0) & 2.68 & 2.65 & \multicolumn{2}{c|}{0.75} & 2$\times$2 & 2.26 \\[2mm]
& {\is} & 2.46 (5.5) & 2.35 (13.0) & 2.57 & 2.51 & \multicolumn{2}{c|}{1.46} & 2$\times$2 & 2.01 \\[2mm]
& {\ib} & 2.51 (3.6) & 2.29 (17.1) & 2.54 & 2.49 & \multicolumn{2}{c|}{2.08} & 2$\times$2 & 1.58 \\[2mm]
& $\,\;{\ih}^{*}$ & \multicolumn{2}{c|}{2.45} & \multicolumn{2}{c|}{2.92} & \multicolumn{2}{c|}{0.62} & 2$\times$2 & 1.39 \\[1mm]
\hline &&&&&&&&& \\[-1mm] 1/2 
& {\cc} & \multicolumn{2}{c|}{2.54} & \multicolumn{2}{c|}{2.68} & \multicolumn{2}{c|}{0.82} & 2$\times$1 & 2.20 \\[2mm]
& {\kk} & \multicolumn{2}{c|}{2.43} & \multicolumn{2}{c|}{2.59} & \multicolumn{2}{c|}{1.52} & 2$\times$2 & 2.19 \\[2mm]
& $\,\;${\ks} & \multicolumn{2}{c|}{2.44} & \multicolumn{2}{c|}{2.55} & \multicolumn{2}{c|}{1.54} & 2$\times$2 & 2.14 \\[2mm]
& {\ph} & 2.87 & 2.91 & 2.66 & 2.93 & 1.32 & 0.57 & 2$\times$(2) & 2.01 \\[2mm]
& {\pp} & \multicolumn{2}{c|}{2.52} & \multicolumn{2}{c|}{2.68} & \multicolumn{2}{c|}{1.39} & 2$\times$1 & 1.75 \\[2mm]
& {\bb} & \multicolumn{2}{c|}{2.55} & \multicolumn{2}{c|}{2.53} & \multicolumn{2}{c|}{2.18} & 2$\times$1 & 1.52 \\[1mm]
\hline &&&&&&&&& \\[-1mm] 1
& {\kk}-{\kk} & \multicolumn{2}{c|}{2.38} & \multicolumn{2}{c|}{2.73} & \multicolumn{2}{c|}{1.49} & 2$\times$1 & 1.98 \\[2mm]
& {\cc}-{\kk} & \multicolumn{2}{c|}{2.45} & 2.75 & 2.79 & 0.81 & 2.00 & 2$\times$2 & 1.93 \\[2mm]
& {\pp}-{\hh} & \multicolumn{2}{c|}{ -- } & 2.65 & 2.87 & \multicolumn{2}{c|}{0.53} & (2)$\times$1 & 1.91 \\[2mm]
& {\cc}-{\pp} & \multicolumn{2}{c|}{2.50} & 2.65 & 2.76 & 0.60 & 1.56 & 2$\times$1 & 1.74 \\[2mm]
& {\cc}-{\bb} & \multicolumn{2}{c|}{2.58} & 2.59 & 2.69 & 0.59 & 2.36 & 2$\times$1 & 1.59 \\[2mm]
& {\pp}-{\bb} & \multicolumn{2}{c|}{2.56} & 2.65 & 3.59 & 1.31 & 3.35 & 2$\times$1 & 1.46 \\[1mm]
\hline\hline
\end{tabular}
\end{minipage}
\end{table*}

\subsection{1/2 ML Coverage}

For the case of 1/2 ML coverage, (i.e., 2 Mg atoms per 2$\times$2 unit cell),
we have considered the adsorption at combinations of pairs of the five
different sites resulting in 15 different cases among which the most favorable
case was found to be the {\cc} sites in our calculations. The structure then
possesses 2$\times$1 reconstruction with symmetric dimers stretched to 2.54
{\AA}, shown in Fig. 2b.

\begin{figure}[bp]
\epsfig{file=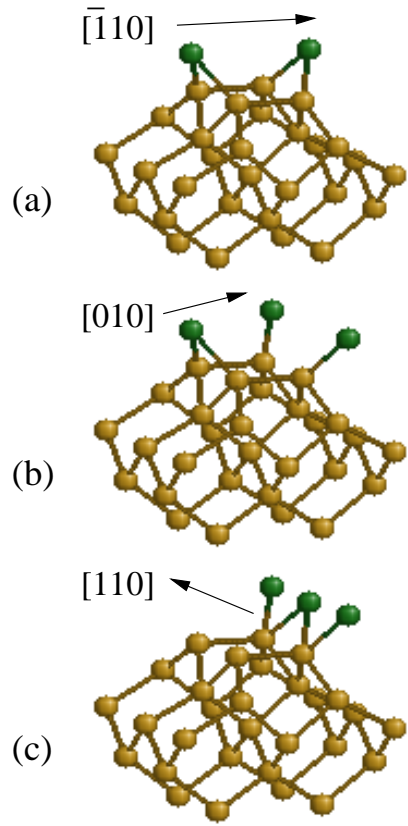,width=6cm,clip=true}
\caption{Adsorption of two Mg atoms on {\is} sites for half coverage: (a) {\kk}
stable, (b) {\ks} stable and (c) {\kk}$^{\prime\prime}$ unstable configurations
\label{figure3}}
\end{figure}

Even though {\is} was stable for the low 1/4 ML coverage, we found that {\kk}
is not always stable for 1/2 ML coverage. Out of four {\is} sites in a
2$\times$2 surface unit cell there are three inequivalent {\kk} geometries as
shown in Fig. 3. For instance, it is unstable if the other adatom occupies the
next {\is} site along the [110], i.e., dimer row direction, as shown in Fig.
3c. However, if the second adatom were added on the other {\is} site along
[$\bar{1}10$], i.e., the direction of dimers, as seen in Fig. 3a, this
2$\times$2 configuration will be stable and it will have an adsorption energy
almost equal to that of the {\cc} case, differing by only 0.01 eV per adatom.
The last case {\ks}, with the second {\is} site is along the [010] direction as
shown in Fig. 3b, is also stable with an adsorption energy of 2.14 eV per
adatom.

Although neither of {\ip} and {\ih} sites are stable at 1/4 ML coverage,
combination of them at 1/2 ML coverage, i.e., the {\ph} case with one atom
occupying a pedestal site while the other occupying the neighboring {\ih} site
was found to be stable and it is more favorable than the occupation of {\pp}
sites that is also stable. In the case of {\pp} adsorption the adsorbed atoms
are four-fold coordinated with the neighboring Si dimer atoms with a bond
length of 2.68 {\AA} and with symmetrical dimers stretched to a length of 2.52
{\AA}. The case of adsorption on {\bb} was also found to be stable with
symmetrical dimers but with an adsorption energy less than the {\cc} case by
0.68 eV per adatom.

The other combinations were found to be not stable and the adatom migrates to
any one of the above stable configuration sites.

In all of the above stable configurations we have found that the surface dimers
become symmetric (e.g., Fig. 2b) in contrast to the 1/4 ML adsorption case
discussed before. The dimer lengths for all the stable adsorption sites are in
the range 2.43--2.55 {\AA} except for the case of {\ph} adsorption where the
two dimers were stretched to rather long values of 2.87 and 2.91 {\AA},
respectively. This asymmetry makes it slightly a 2$\times$2 rather than a
2$\times$1 like most other cases in 1/2 ML coverage with the exception of the
{\kk} case which is naturally a 2$\times$2. As to the Mg--Si backbonds, they
are in the range of 2.53--2.93 {\AA}. In Table~\ref{table1} the d$_{\rm Mg-Si}$
values listed for the {\ph} case are for the Mg atom on {\ip} and {\ih} sites
bonded to dimer Si atoms, respectively. Similarly, the d$_\perp$ values
correspond to the heights measured from Si dimers of Mg adatoms on {\ip} and
{\ih} sites, respectively.

In addition to the pathway $b\rightarrow s\rightarrow c$, we have also
$b\rightarrow p\rightarrow s\leftrightarrow h\leftrightarrow c$ pathway in 1/2
ML coverage, where the backward migrations in energy take place among sites in
the trough. Another rule brought by the double occupancy in this coverage is
that the adatoms form chains along the dimer rows or along the dimer bonds. The
{\ks}-case looks like an exception to this, however, it is not considered to be
a chain to first order. Any possible unstable combination $\alpha$-$\beta$
phase will diffuse into one of the stable phases listed in Table~\ref{table1}
by the adatom on either one of $\alpha$ or $\beta$ sites migrating to a new
site according to the above rules. In the case of {\it b-c} phase, since each
Si atom of all dimers are bonded to two Mg atoms it is not stable and
consequently, both Mg atoms at {\ib} and {\ic} sites migrate to their
corresponding neighbors {\ip} and {\ih} sites, respectively, to reach the
stable {\ph} phase. We should emphasize here that this conclusion is not only
based on the total energy calculations but also on the comparisons of the
Hellman-Feynman forces and the adatom behavior in the other less stable cases.

\subsection{1 ML Coverage}

For the full coverage adsorption, we have found six most probable
configurations that are also consistent with our findings for the half coverage
case. They are listed in Table~\ref{table1} in the order of their adsorption
energies.

The adsorption on {\kk}-{\kk} was found to be the most favorable one with an
adsorption energy of 1.98 eV per adatom. In this model we have four Mg atoms
occupying all four {\is} sites forming Mg lines that run along the dimer rows.
The surface symmetry of this structure is 2$\times$1 with dimers shortened to
2.38 {\AA} as compared to 2.43 {\AA} of the {\kk} half coverage case. The Mg-Si
backbond of length 2.73 {\AA} compares well with the Mg-Si bond in Mg$_2$Si.

The next stable case that has less adsorption energy (by only 0.05 eV per
adatom) is the {\cc}-{\kk} in which one has two Mg adatoms on {\kk} along the
[$\bar{1}10$] direction, while the other two Mg adatoms occupy the {\cc} sites
along the [$110$] direction as seen in Fig. 4b. In this case the reconstructed
surface structure possesses a 2$\times$2 symmetry with dimers of length 2.45
\AA. The Mg-Si backbonds being in the range of 2.75--2.79 {\AA} also compare
well with the Mg-Si bonds in silicates.

The {\pp-\hh} combination, in which every hallow and pedestal sites are
occupied, was also found to be stable with an adsorption energy differing from
the {\kk}-{\kk} case by 0.07 eV per adatom. This phase is very interesting in
its symmetry properties. Upon relaxation the dimers are removed and it
approached to an almost 1$\times$1 phase where {\ip} and {\ih} become the same.
The deviation from 1$\times$1 is very small and is due to a slight shift (0.15
{\AA}) of only the ex-dimer-member Si atoms in the [$\bar{1}$10] direction
opposite to one another, so that the Mg adatom will only bond to two Si atoms
instead of four. This symmetry breaking due to charge transfer from Mg to
surface Si atoms causes the zigzag bond picture shown in Fig. 4c.

\begin{figure}[htbp]
\epsfig{file=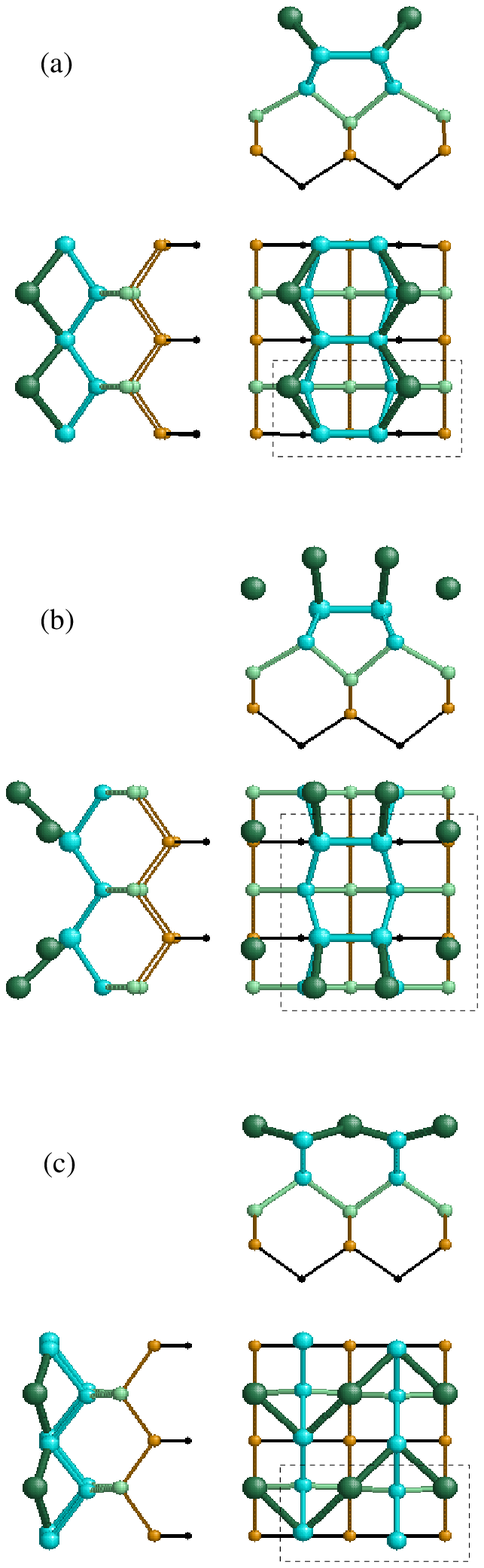,width=7.4cm,clip=true} \caption{Three views of the minimum
energy structures (after relaxation) for (a) Mg/Si(001)-(2$\times$1)-{\kk-\kk},
(b) Mg/Si(001)-(2$\times$2)-{\cc-\kk} and (c) Mg/Si(001)-(2$\times$1)-{\pp-\hh}
cases for 1 ML coverage\label{figure4}}
\end{figure}

Next combination is the {\cc}-{\pp} case where the adsorbed atom in {\ip} site
is four-fold coordinated with the neighboring Si dimer atoms with a bond length
of 2.65 {\AA} while the other adatom in cave site is two-fold coordinated with
a bond length of 2.76 {\AA} giving rise to an adsorption energy of 1.74 eV per
adatom. This structure possesses 2$\times$1 symmetry with a dimer length of
2.50 {\AA}.

The combination {\cc-\bb} was also found stable with adsorption energy equal to
1.59 eV per adatom. In this model we have Mg atoms occupying all the bridge and
cave sites, forming Mg lines that run orthogonal to the dimer rows and
resulting in a 2$\times$1 symmetry. The dimer length was found to be 2.58 \AA.
The difference in heights of the Mg adatoms on {\ic} and {\ib} sites measured
from the silicon dimers corresponding to 1.77 {\AA} undulates the Mg chains
along the dimer direction.

The configuration {\pp-\bb} can be described as double half monolayer where two
Mg atoms occupy two pedestal sites ($d_\perp$=1.31 \AA) while the other two Mg
adatoms rise further up to a height of 3.35 {\AA} measured from the center of
the dimer. This structure possesses 2$\times$1 symmetry with a dimer length of
2.56 {\AA} and can be described as undulated Mg wires located on the Si dimer
rows, similar but orthogonal to the {\cc-\bb} case.

\section{Summary and Conclusion}

We have performed an ab initio total energy calculation and geometry
optimization for a clean Si(001) surface and that with Mg overlayer of
different coverages on it. For the 1/4 ML coverage we have found two favorable
adsorption sites, namely {\ic} and {\is} with a relative adsorption energy
difference of only 0.25 eV per adatom. Our results agree well with the
room-temperature results of Hutchison et al.\cite{HEN} and of Cho et
al.\cite{Cho} where they have suggested that the most favorable site is the
cave site. Kubo et al.\cite{Kubo} have suggested for the higher temperature
adsorption that the {\is} site is favorable, however, they assumed an extra Si
on the surface bonding to the adatom. Even though Kubo et al.\cite{Kubo}
suggested the possibility of single adatom occupation on a pedestal site at
room temperature, we did not find any evidence to confirm it. Our calculations
show that this site is unstable for a single adatom adsorption. Occupation of
pedestal site has also been ruled out by Cho et al.\cite{Cho} for low coverage.
But we find that the argument of Hutchison et al.\cite{HEN} that this site can
be occupied by multiple atoms is in agreement with our stable {\pp} case for
1/2 ML.

Contrary to the theoretical results\cite{Wang} for Ba adsorption on Si(001), we
found that the Mg adsorption on {\ih} site is metastable, i.e., the adatom
resides there temporarily, altering the dimers into symmetrical ones. However,
the total energy of this case being higher than the adsorption on {\ic} site by
0.85 eV per adatom, eventually, it leaves this saddle point and migrates to the
neighboring cave site. We believe the reason for this disagreement with the Ba
case might be due to the difference in their atomic sizes.

For the case of half monolayer coverage, our results give 2$\times$1
reconstruction with the Mg adatoms on {\cc} sites. The high temperature result
of Kubo et al.,\cite{Kubo} for the same coverage suggested that it contains two
Mg adatoms located nearly on {\cc} sites but distorted by the existence of an
extra Si atom which causes the symmetry to be a 2$\times$2. On the other hand,
their 1/2 ML result for low temperature can be described as two Mg adatoms
confined in every other cave site and separated towards {\ih}-saddle to make it
a 2$\times$2. This corresponds to a distorted version of our {\hh} case.
Starting from both of these geometries we have found them unstable ending up in
{\cc} case with a 2$\times$1 symmetry. Even though we have started with a
2$\times$2 unit cell for each case in our calculations we have ended up with
2$\times$1 reconstruction for the 1/2 ML coverage except for the {\kk} cases,
which are naturally 2$\times$2, and the {\ph} case which is slightly a
2$\times$2 distorted from 2$\times$1. This slight symmetry breaking shows up
itself in 1 ML coverage case of {\pp-\hh} as well.

The bridge site {\bb} was found to be stable in accordance with the theoretical
result of Khoo and Ong\cite{KO} for this coverage, even though it is the least
favorable case.

For the case of full monolayer coverage, we have found that the adsorption
model in which all the Mg atoms occupy all the {\is} sites with 2$\times$1
construction is more favorable than 2$\times$2 reconstruction in which Mg atoms
occupy two shallow and two cave sites having an adsorption energy difference of
only 0.05 eV per adatom. The next favorable case having 0.07 eV per adatom
smaller adsorption energy than the most preferable {\kk-\kk} case is the
{\pp-\hh} configuration with a 2$\times$1 symmetry, but just slightly away from
1$\times$1 geometry as seen in Fig. 4c. This may suggest that more than 1 ML
coverage might lead to a 1$\times$1 phase consistent with the LEED results of
Kim et al.\cite{Kim} who reported that the 1$\times$1 phase corresponds to 2 ML
coverage.

The model of adsorption that can be concluded from our theoretical results is
that for the low coverage, Mg atoms will be adsorbed mostly on {\ic} sites and
less likely on {\is} sites, which continues until the coverage reaches to 1/4
ML. As the coverage exceeds 1/4 ML, the Mg atoms are adsorbed on the {\cc}
sites and on the {\kk} sites along [$\bar{1}10$] with almost equal
probabilities until the surface becomes saturated at 1/2 ML, where it is going
to have domains of {\cc} and {\kk} phases with 2$\times$1 and 2$\times$2
reconstructions, respectively. Further increase in coverage will result into
{\kk-\kk}, {\cc-\kk} and even {\pp-\hh} domains with 2$\times$1, 2$\times$2 and
almost 1$\times$1 reconstructions, respectively.

Using ab initio total energy calculations we have performed detailed
investigation of the atomic geometry of Mg adsorbed Si(001) surface for 1/4,
1/2 and full monolayer coverages. We have investigated the adsorption sites,
the energetics and the reconstruction of Si(001) surface after Mg adsorption.
We have found that our results agree well with the existing experimental
observations.

\begin{acknowledgments}
This work was supported by T{\"U}B\.{I}TAK, The Scientific and Technical
Research Council of Turkey, Grant No. TBAG-2036 (101T058).
\end{acknowledgments}

\bibliography{basename of .bib file}

\end{document}